# Three-Stage Speaker Verification Architecture in Emotional Talking Environments

Ismail Shahin and [*]Ali Bou Nassif

Department of Electrical and Computer Engineering

University of Sharjah

P. O. Box  27272

Sharjah, United Arab Emirates

Tel: (971) 6 5050967

Fax: (971) 6 5050877

E-mail: ismail@sharjah.ac.ae, [*]anassif@sharjah.ac.ae

## Abstract

Speaker verification performance in neutral talking environment is usually high, while it is sharply decreased in emotional talking environments. This performance degradation in emotional environments is due to the problem of mismatch between training in neutral environment while testing in emotional environments. In this work, a three-stage speaker verification architecture has been proposed to enhance speaker verification performance in emotional environments. This architecture is comprised of three cascaded stages: gender identification stage followed by an emotion identification stage followed by a speaker verification stage. The proposed framework has been evaluated on two distinct and independent emotional speech datasets: in-house dataset and "Emotional Prosody Speech and Transcripts" dataset. Our results show that speaker verification based on both gender information and emotion information is superior to each of speaker verification based on gender information only, emotion information only, and neither gender information nor emotion information. The attained average speaker verification performance based on the proposed framework is very alike to that attained in subjective assessment by human listeners.

## 1.    Introduction

"Speaker identification and speaker verification (authentication)" are two main branches of speaker recognition. "Speaker identification" is the process of identifying the unknown speaker from a set of known speakers, while "speaker verification" is the process of accepting or rejecting the claimed speaker. This branch is considered as a true-or-false binary decision problem. "Speaker identification" can be utilized in investigating criminals to decide the suspects who produced the voice captured during the crime. "Speaker verification" technologies have wide range of applications such as: biometric person authentication, speaker verification for surveillance, forensic speaker recognition, and security applications including credit card transactions, computer access control, monitoring people, telephone voice authentication for long distance calling or banking access [1].

Speaker recognition comes in two forms in terms of spoken text: "text-dependent" and "text-independent". In "text-dependent", the same text is uttered in both training and testing phases, while in "text-independent", there is no restriction of voice sample in the training and testing phases.

In this work, we address the issue of improving "speaker verification performance in emotional environments" based on proposing, applying, and evaluating a three-stage speaker verification architecture that is consists of three cascaded stages: "gender identification stage" followed by an "emotion identification stage" followed by a "speaker verification stage".

## 2.     Prior Work

Speaker verification performs almost ideally in neutral talking environment, while it performs poorly in emotional talking environments. There are many studies that study "speaker verification in neutral environment" [2-6], while few studies spotlight on "speaker verification in emotional environments" [7-11].

Speaker recognition has been an attractive research field in the last few decades, which still yields a number of challenging problems. One of the most challenging problems that faces speaker recognition researchers is the low performance in emotional environments [7-10]. Emotion-based speaker recognition is one of the central research fields in the human-computer interaction or affective computing area [12], [13], [14]. The main target of intelligent human-machine interaction is to empower computers with the affective computing capability so that machines can recognize users in intelligent services.

There are many research [2-6] that study speaker verification in neutral environments. The authors of [2] aimed in one of their work at addressing the long-term speaker variability problem in the feature domain in which they extracted more exact speaker-specific and time-insensitive information. They tried to recognize frequency bands that expose greater discrimination for speaker-specific data and lower sensitivity with respect to diverse sessions. Their strategy was based on the $F$-ratio criterion to decide the whole discrimination-sensitivity of frequency bands by including both the session-specific variability data and the speaker-specific information [2]. The authors of [3] proposed extracting local session variability vectors on distinct phonetic classes from the utterances instead of estimating the session variability across the overall utterance as i-vector does. Based on the deep neural network (DNN), the posteriors trained for phone state categorization, local vectors express the session variability contained in specific phonetic content. Their experiments demonstrated that the content-aware local vectors are superior to the DNN i-vectors in the trials where short utterances are involved [3]. The authors of [4] focused on the issues associated with language and speaker recognition, studying prosodic features extracted from speech signals. Their proposed method was tested using the "National Institute of Standards and Technology (NIST) language recognition evaluation 2003" and the extended data task of "NIST speaker recognition evaluation 2003 for language and speaker recognition", respectively. The authors of [5] described the main components of "MIT Lincoln Laboratory's Gaussian Mixture Model (GMM)-based speaker verification system in a neutral environment". The authors of [6] directed their work on "text-dependent speaker verification systems" in a such environment. In their proposed framework, they utilized "suprasegmental and source features, in addition to spectral features" to authenticate speakers. The combination of "suprasegmental, source, and spectral features" considerably improves speaker verification performance [6].

In contrast, there are less number of studies [7-11] that study the problem of "speaker verification in emotional environments". The authors of [7] presented investigations into the effectiveness of the state-of-the-art speaker verification techniques: "Gaussian Mixture Model-Universal Background Model and Gaussian Mixture Model-Support Vector Machine (GMM-UBM and GMM-SVM)" in mismatched noise conditions. The authors of [8] tested whether speaker verification algorithms that are trained in emotional environments give better performance when implemented to speech samples achieved under stressful or emotional conditions than those trained in a neutral environment only. Their conclusion is that training of speaker verification algorithms on a wider range of speech samples, including stressful and emotional talking conditions, rather than the neutral talking condition, is a promising method to improve speaker authentication performance [8]. The author of [9] proposed, applied, and evaluated a two-stage approach for speaker verification systems in emotional environments based completely on "Hidden Markov Models (HMMs)". He examined the proposed approach using a collected speech dataset and obtained 84.1% as a speaker verification performance. The authors of [10] studied the impact of emotion on the performance of a "Gaussian Mixture Model-Universal Background Model (GMM-UBM) based speaker verification system" in such environments. In their study, they proposed an emotion-dependent score normalization framework for speaker verification on emotional speech. They reported an "average speaker verification performance" of 88.5% [10]. In [11], the author focused on employing and evaluating a two-stage method to authenticate the claimed speaker in emotional environments. His method is made up of two recognizers which are combined and integrated into one recognizer using both "HMMs and Suprasegmental Hidden Markov Models (SPHMMs)" as classifiers. The two recognizers are: an "emotion identification recognizer" followed by a "speaker verification recognizer". He attained average Equal Error Rate (EER) of 7.75% and 8.17% using a collected dataset and "Emotional Prosody Speech and Transcripts (EPST)" dataset, respectively.

The main contribution of the present work is to further enhance speaker verification performance compared to that based on the two-stage approach [11] by employing and testing a three-stage speaker verification architecture to verify the claimed speaker in emotional environments. This architecture is comprised of three recognizers that are combined and integrated into one recognizer using both "HMMs and SPHMMs" as classifiers. The three recognizers are: gender identification recognizer followed by an emotion identification recognizer followed by a speaker verification recognizer. Specifically, our current work focuses on improving the performance of text-independent, gender-dependent, and emotion-dependent speaker verification system in such environments. This work deals with inter-session variability caused by distinct emotional states of the claimed speaker. Based on the proposed framework, the claimed speaker should be registered in advance in the test set (closed set). Our present work is different from two of our preceding studies [11, 15]. In [11], we focused on verifying the claimed speaker based on a two-stage framework (speaker verification stage preceded by an emotion identification stage) in emotional environments. In [15], we focused on identifying speakers in emotional environments based on a three-stage framework (gender identification phase followed by an emotion identification phase followed by a speaker identification phase).

The proposed architecture in the current research centers on enhancing low speaker verification performance in emotional environments based on employing both of gender and emotion cues. This work is a continuation to one of our prior work [11] which was devoted to proposing, applying, and assessing a two-stage method to authenticate speakers in emotional environments based on "SPHMMs and HMMs" as classifiers. Moreover, seven extensive experiments have been performed in the present research to assess the proposed three-stage architecture.

Specifically, in this paper, we raise the following research questions:

**RQ1**: Does the three-stage framework increase the performance of speaker verification in emotional environments in comparison to:

    **RQ1.1**  A single-stage framework?

    **RQ1.2**  Emotion independent two-stage framework?

    **RQ1.3**  Gender independent two-stage framework?

**RQ2**: As classifiers, which is more superior on the three-stage speaker verification, "HMMs or SPHMMs"?

The rest of the work is structured as follows: Section 3 covers the basics of SPHMMs. Section 4 describes the two speech datasets used to assess the proposed architecture and the extraction of features. The three-stage framework and the experiments are discussed in Section 5. Section 6 presents decision threshold. The attained results in the current work and their discussion are demonstrated in Section 7. Finally, Section 8 gives the concluding remarks of this work.

### 3.    Basics of Suprasegmental Hidden Markov Models

"SPHMMs" were applied and assessed by Shahin in many occasions: speaker identification in each of emotional and shouted environments [15,16,17], speaker verification in emotional environments [11], and emotion recognition [18,19]. In these studies, "SPHMMs" have shown to be superior models over "HMMs". This is because "SPHMMs" have the capability to summarize some states of "HMMs" into a new state named "suprasegmental state". "Suprasegmental state" has the ability to look at the observation sequence through a bigger window. This state allows observations at rates proper for the case of modeling emotional and stressful signals. Prosodic data cannot be perceived at a rate that is utilized for "acoustic modeling". The "prosodic features" of a unit of emotional and stressful signals are coined "suprasegmental features" since they have the effect on all the segments of the unit signal. Prosodic events at the levels of "phone, syllable, word, and utterance" are expressed utilizing "suprasegmental states", while acoustic events are modeled using "conventional hidden Markov states".

Polzin and Waibel [20] combined and integrated prosodic data with acoustic data within HMMs as given by,

$$\text{``}log\ \text{P}\left(\lambda^{v}, \Psi^{v}\,|\,\text{O}\right) = (1 - \alpha).\ log\ \text{P}\left(\lambda^{v}\,|\,\text{O}\right) + \alpha.\ log\ \text{P}\left(\Psi^{v}\,|\,\text{O}\right) \tag{1}$$

where $\alpha$ is a weighting factor. When:

$$
\begin{cases}
0.5 > \alpha > 0 & \text{biased towards acoustic model} \\
1 > \alpha > 0.5 & \text{biased towards prosodic model} \\
\alpha = 0 & \text{biased completely towards acoustic model and} \\
& \text{no effect of prosodic model} \\
\alpha = 0.5 & \text{not biased towards any model} \\
\alpha = 1 & \text{biased completely towards prosodic model and} \\
& \text{no impact of acoustic model}
\end{cases} \tag{2}
$$

$\lambda^{v}$ is the $v^{\text{th}}$ acoustic model, $\Psi^{v}$ is the $v^{\text{th}}$ SPHMM model, $O$ is the observation vector of an utterance, $P\left(\lambda^{v}\,|\,O\right)$ is the probability of the $v^{\text{th}}$ HMM model given the observation vector $O$, and $P\left(\Psi^{v}\,|\,O\right)$ is the probability of the $v^{\text{th}}$ SPHMM model given the observation vector $O$".

Eq. (1) demonstrates that departing "a suprasegmental state requires summing the log probability of this suprasegmental state given the relevant suprasegmental observations within the emotional/stressful signal to the log probability of the current acoustic model given the particular acoustic observations within the signal. Additional information about SPHMMs can be attained from the references [16,17,18,19]".

## 4. Speech Datasets and Extraction of Features

In the present research, our proposed three-stage speaker verification architecture has been evaluated on two diverse and independent emotional datasets: in-house dataset and "Emotional Prosody Speech and Transcripts (EPST) Dataset".

### 4.1 In-House Dataset

Twenty men and twenty women untrained adult (with ages spanning between 18 years and 55 years) native speakers of American English generated the collected speech dataset in this work. The untrained forty speakers were chosen to spontaneously utter eight sentences and to keep away from overstressed expressions. Each speaker was asked to utter eight sentences where each sentence was spoken nine times under each of "neutral, anger, sadness, happiness, disgust, and fear emotions". The eight sentences were carefully selected to be unbiased towards any emotion. The sentences are:

1) *"He works five days a week.*
2) *The sun is shining.*
3) *The weather is fair.*
4) *The students study hard.*
5) *Assistant professors are looking for promotion.*
6) *University of Sharjah.*
7) *Electrical and Computer Engineering Department.*
8) *He has two sons and two daughters."*

The "first four sentences" of this dataset were utilized in the "training phase"; on the other hand, the "last four sentences" were utilized in the "evaluation phase" (text-independent problem). The captured speech dataset was collected in an uncontaminated environment by a "speech acquisition board using a 16-bit linear coding A/D converter and sampled at a sampling rate of 16 kHz". This dataset is a wideband 16-bit per sample linear data. A pre-emphasizer was applied to the speech signal samples. Then, these signals were sliced into slices (frames) of 16 ms each with 9 ms intersection between adjacent slices. The emphasized speech signals were applied every 5 ms to a 30 ms Hamming window.

## 4.2 "Emotional Prosody Speech and Transcripts (EPST) Dataset"

EPST dataset was introduced by "Linguistic Data Consortium (LDC)" [21]. This dataset was generated by eight professional speakers ("three actors and five actresses") generating a sequence of "semantically neutral utterances made up of dates and numbers" spoken in fifteen distinct emotions including the neutral state. Only six emotions ("neutral, hot anger, sadness, happiness, disgust, and panic") were utilized in this study. Using this dataset, only four utterances were utilized in the "training phase", while another different four utterances were utilized in the "evaluation phase" (text-independent problem).

## 4.3 Extraction of Features

"Mel-Frequency Cepstral Coefficients (MFCCs)" have been utilized as the extracted features that characterize the phonetic content of speech signals in the two datasets. These coefficients have been largely used in many work in the areas of speech recognition [22], [23], speaker recognition [11], [15], [24], [25], and emotion recognition [17], [26], [27]. This is because these coefficients have proven to be superior to other coefficients in these areas and because they give a high-level estimation of human auditory perception [25], [28].

The vast majority of studies [29], [30], [31] that have been conducted in the last few decades in the areas of speech recognition, speaker recognition, and emotion recognition on "HMMs" have been implemented using "Left-to-Right Hidden Markov Models (LTRHMMs)" since phonemes firmly follow left-to-right sequence. In the present research, "Left-to-Right Suprasegmental Hidden Markov Models (LTRSPHHMs)" have been derived from "LTRHMMs". Fig. 1 illustrates an example of a basic structure of "LTRSPHMMs" that has been obtained from "LTRHMMs". In this figure, "$q_1$, $q_2$, ..., $q_6$" are considered "conventional hidden Markov states". $p_1$ is a "suprasegmental state" that is made up of "$q_1$, $q_2$, and $q_3$". $p_2$ is a "suprasegmental state" that is composed of "$q_4$, $q_5$, and $q_6$". $p_3$ is a "suprasegmental state" that is comprised of "$p_1$ and $p_2$". The transition probability between the $i^{\text{th}}$ conventional hidden Markov state and the $j^{\text{th}}$ conventional hidden Markov state is symbolized by $a_{ij}$. The transition probability between the $i^{\text{th}}$ suprasegmental state and the $j^{\text{th}}$ suprasegmental state is denoted by $b_{ij}$.

In the present work, the number of "conventional states" of "LTRHMMs", $N$, is six. The number of mixture components, $M$, is ten per state, with a continuous mixture observation density is chosen for such models. The number of "suprasegmental states" in "LTRSPHMMs" is two. Consequently, each three "conventional states of LTRHMMs" are condensed into one "suprasegmental state". "The transition matrix, $A$," of such a structure can be defined in terms of the "positive coefficients $b_{ij}$ as,

$$A = \begin{bmatrix} b_{11} & b_{12} \\ 0 & b_{22} \end{bmatrix},$$

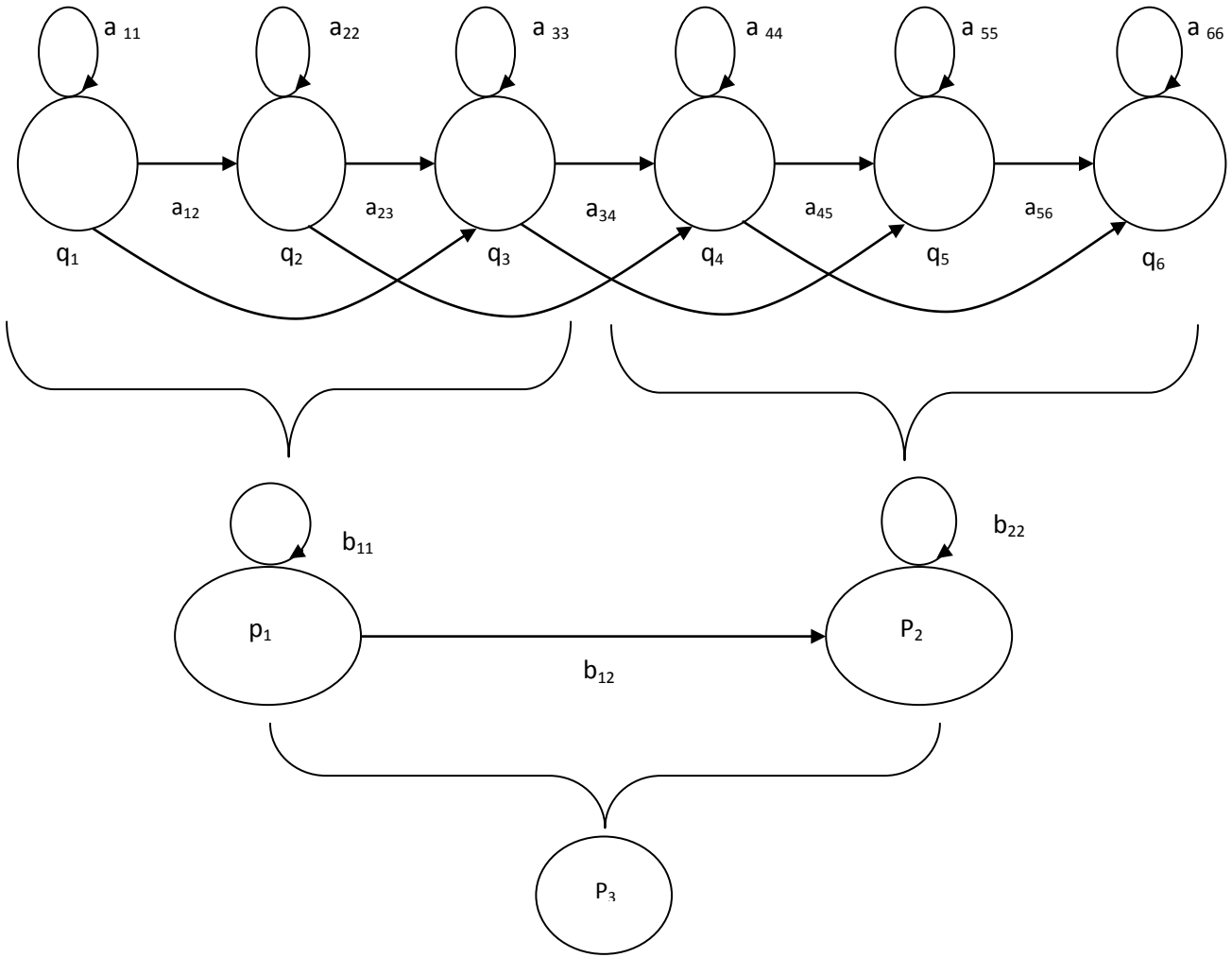

**Fig. 1.** Basic structure of LTRSPHMMs

## 5. Three-Stage Speaker Verification Architecture and the Experiments

Given *n* speakers for each gender where every speaker emotionally talks in *m* emotions, the overall proposed architecture consists of three sequential stages as shown in Fig. 2. Fig. 2 demonstrates that the proposed architecture is comprised of three cascaded recognizers that integrate and combine gender identifier followed by emotion identifier followed by speaker verifier into one architecture. The three stages are:

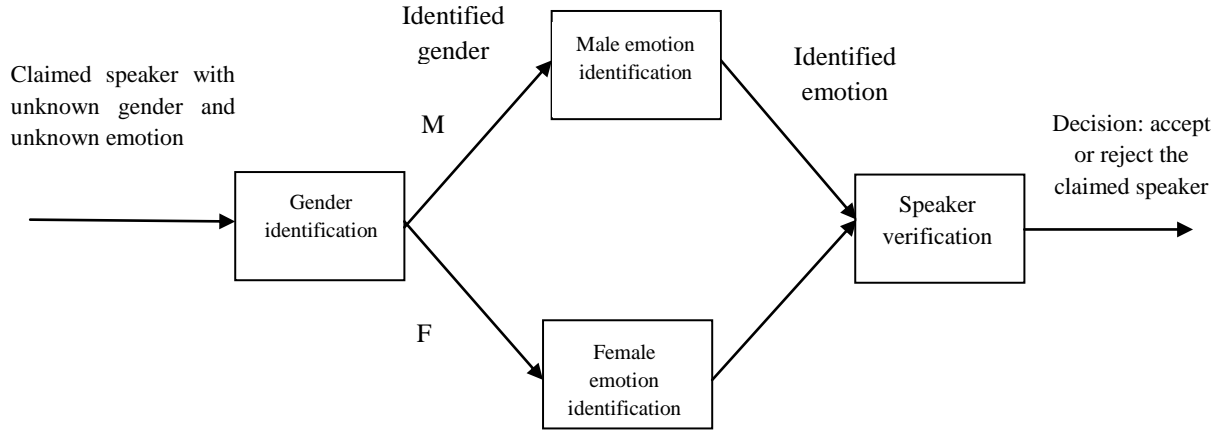

**Fig. 2.** Block diagram of the overall proposed three-stage speaker verification architecture

## 5.1 Stage 1: "Gender Identification Stage"

The first stage of the overall three-stage architecture is to recognize the gender of the claimed speaker in order to make the output of this stage gender-dependent. Typically, automatic gender identification step yields high performance without much work because the output of this stage is the claimed speaker either a male or a female. So, gender identification problem is a binary classification which is normally not a very challenging step.

In the current stage, two probabilities for each utterance are calculated based on HMMs and the maximum probability is chosen as the recognized gender as shown in the given formula,

$$\text{G}^* = \arg \max_{2 \geq g \geq 1} \left\{ \text{P}\left( \text{O} \middle| \Gamma^g \right) \right\} \qquad (3)$$

where "$G^*$ is the pointer of the recognized gender (either $M$ or $F$), $\Gamma^g$ is the $g^{\text{th}}$ HMM gender model, and $P\left(O \middle| \Gamma^g\right)$ is the probability of the observation sequence $O$ that corresponds to the unknown gender of the claimed speaker given the $g^{\text{th}}$ HMM gender model".

In the "training session" of this stage, "HMM male gender model" has been constructed using the "twenty male speakers" generating all the "first four sentences" under all the emotions, while "HMM female gender model" has been derived using the "twenty female speakers" producing all the "first four sentences" under all the emotions. The total number of utterances used to build each "HMM gender model" is 4320 (20 speakers $\times$ 4 sentences $\times$ 9 utterances/sentence $\times$ 6 emotions).

## 5.2 Stage 2: "Emotion Identification Stage"

Given that the gender of the claimed speaker was recognized in the preceding stage, the goal of this stage is to recognize the unknown emotion of the claimed speaker who is speaking emotionally. This stage is named "gender-specific emotion identification". In this stage, there are $m$ probabilities per gender that are calculated using SPHMMs. The highest probability is selected as the recognized emotion per gender as shown in the given formula,

$$E^* = \arg \max_{m \geq e \geq 1} \left\{ P\left(O \middle| G^*, \lambda_E^e, \Psi_E^e\right) \right\} \tag{4}$$

where "$E^*$ is the index of the identified emotion, $\left(\lambda_E^e, \psi_E^e\right)$ is the $e^{\text{th}}$ SPHMM emotion model, and $P\left(O \middle| G^*, \lambda_E^e, \Psi_E^e\right)$ is the probability of the observation sequence $O$ that belongs to the unknown emotion given the identified gender and the $e^{\text{th}}$ SPHMM emotion model".

In the emotion identification stage, the "$e^{th}$ SPHMM emotion model" $\left( \lambda_E^e, \Psi_E^e \right)$ per gender has been obtained in the "training phase" for every emotion using the "twenty speakers" per gender generating all the "first four sentences with a replication of nine utterances/sentence". The overall number of utterances utilized to derive every "SPHMM emotion model" for each gender is 720 (20 speakers × 4 sentences × 9 utterances/sentence). The "training phase" of "SPHMMs" is very alike to the "training phase of the conventional HMMs". "Suprasegmental models" are trained on top of "acoustic models of HMMs" in the training phase of SPHMMs. This stage is shown in a block diagram of Fig. 3.

## 5.3 Stage 3: "Speaker Verification Stage"

The last stage of the overall three-stage framework is to verify the speaker identity based on HMMs given that both of his/her gender and emotion were identified in the previous two stages ("gender-specific and emotion-specific speaker verification problem") as shown in the following formula,

$$\Lambda(\text{O}) = log\left[ \text{P}\left( \text{O} \middle| \text{E}^*, G^* \right) \right] - log\left[ \text{P}\left( \text{O} \middle| \overline{\text{E}}^*, G^* \right) \right] - log\left[ \text{P}\left( \text{O} \middle| \overline{\text{E}}^*, \overline{G}^* \right) \right] \tag{5}$$

where "$\Lambda(O)$ is the log-likelihood ratio in the $log$ domain, $P\left( O \middle| E^*, G^* \right)$ is the probability of the observation sequence $O$ that belongs to the claimed speaker given the true recognized emotion and the true recognized gender, $P\left( O \middle| \overline{E}^*, G^* \right)$ is the probability of the observation sequence $O$ that corresponds to the claimed speaker given the incorrect recognized emotion and the true recognized gender, and $P\left( O \middle| \overline{E}^*, \overline{G}^* \right)$ is the probability of the observation sequence $O$ that belongs to the claimed speaker given the wrong recognized emotion and the incorrect recognized gender". Eq. (5) shows that the likelihood ratio is computed among model trained using data from recognized gender, recognized emotion, and claimed speaker.

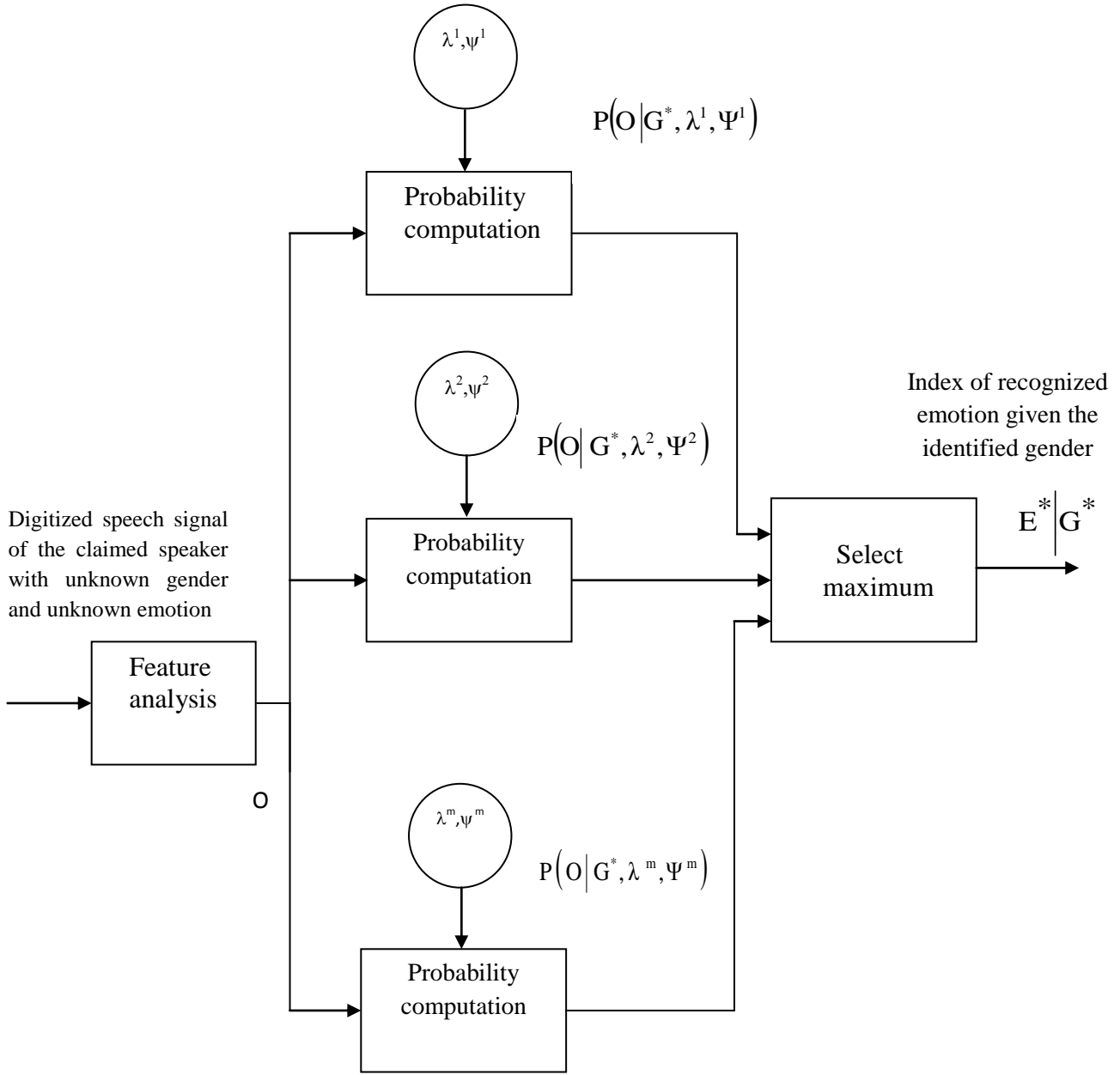

**Fig. 3.** Block diagram of stage 2 of the whole proposed three-stage architecture

The probability of the observation sequence $O$ that belongs to the claimed speaker given the true recognized emotion and the true recognized gender can be calculated as [32],

$$log\ P\left(O\left|\ E^{*},G^{*}\right.\right)=\frac{1}{T}\sum_{t=1}^{T}log\ P\left(o_{t}\left|\ E^{*},G^{*}\right.\right)$$ (6)

where, $O = o_1 o_2 ... o_t ... o_T$.

The probability of the observation sequence $O$ that corresponds to the claimed speaker given the wrong recognized emotion and the true recognized gender can be obtained using a set of $B$ imposter emotion models: $\left\{\overline{E}_{1}^{*},\overline{E}_{2}^{*},...,\overline{E}_{B}^{*}\right\}$ as,

$$log\ P\left(O\left|\ \overline{E}^{*},G^{*}\right.\right)=\left\{\frac{1}{B}\sum_{b=1}^{B}log\left[P\left(O\left|\ \overline{E}_{b}^{*},G^{*}\right.\right)\right]\right\}$$ (7)

where $P\left(O\left|\ \overline{E}_{b}^{*},G^{*}\right.\right)$ can be calculated using Eq. (6). In the current work, the value of $B$ is equal to $6 - 1 = 5$ emotions.

The probability of the observation sequence $O$ that corresponds to the claimed speaker given the incorrect recognized emotion and the wrong recognized gender can be determined using the same set of $B$ imposter emotion models as,

$$log\ P\left(O\left|\ \overline{E}^{*},\overline{G}^{*}\right.\right)=\left\{\frac{1}{B}\sum_{b=1}^{B}log\left[P\left(O\left|\ \overline{E}_{b}^{*},\overline{G}^{*}\right.\right)\right]\right\}$$ (8)

where $P\left(O\left|\ \overline{E}_{b}^{*},\overline{G}^{*}\right.\right)$ can be calculated using Eq. (6). A demonstration of this stage is given in a block diagram of Fig. 4.

In the evaluation phase, every one of the forty speakers utilized nine utterances per sentence of the "last four sentences" (text-independent) under every emotion. The entire number of utterances utilized in this phase is 8640 (40 speakers × 4 sentences × 9 utterances / sentence × 6 emotions). Seventeen speakers per gender have been utilized as claimants and the remaining speakers have been utilized as imposters in this work.

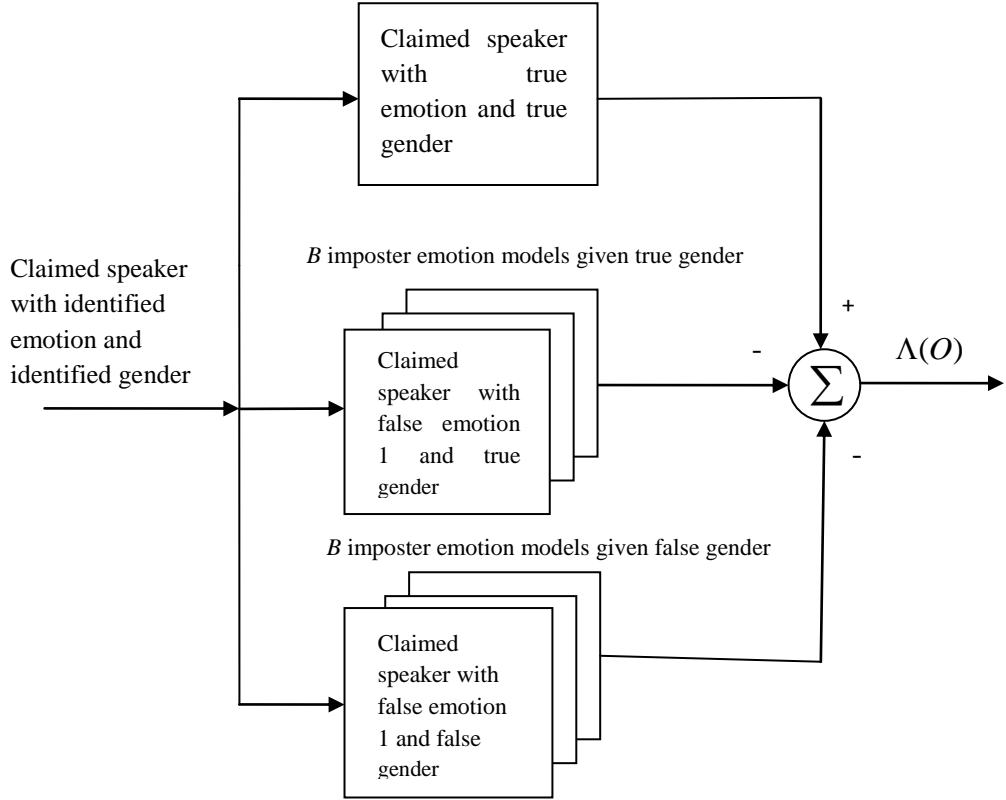

**Fig. 4.** Block diagram of stage 3 of the overall proposed three-stage architecture

## 6.    Decision Threshold

In a speaker verification problem, two types of error can occur: "false rejection" (miss probability) and "false acceptance" (false alarm probability). When a correct identity claim is rejected, it is named a "false rejection error"; in contrast, when the identity claim from an imposter is accepted, it is termed "a false acceptance".

Speaker verification problem, based on emotion identification given the identified gender, involves making a binary decision based on two hypotheses: "hypothesis $H_0$" if the claimed speaker corresponds to a true emotion given the identified gender or "hypothesis $H_1$" if the claimed speaker comes from a false emotion given the identified gender.

The "log-likelihood ratio in the log domain" can be defined as,

$$\Lambda(O) = log\left[ P\left(O\big|\lambda_C, \Psi_C, \ \Gamma_C\right)\right] - log\left[ P\left(O\big|\lambda_{\overline{C}}, \Psi_{\overline{C}}, \Gamma_C\right)\right] - log\left[ P\left(O\big|\lambda_{\overline{C}}, \Psi_{\overline{C}}, \Gamma_{\overline{C}}\right)\right] \tag{9}$$

where "$O$ is the observation sequence of the claimed speaker, $(\lambda_C, \Psi_C)$ is the SPHMM claimant emotion model, $\Gamma_C$ is the HMM claimant gender model, $P(O|\lambda_c, \Psi_c, \Gamma_c)$ is the probability that the claimed speaker belongs to a true identified emotion and a true identified gender, $(\lambda_{\overline{c}}, \Psi_{\overline{c}})$ is the SPHMM imposter emotion model, $\Gamma_{\overline{C}}$ is the HMM imposter gender model, $P(O|\lambda_{\overline{c}}, \Psi_{\overline{c}}, \Gamma_c)$ is the probability that the claimed speaker comes from a false identified emotion and a true identified gender, and $P(O|\lambda_{\overline{c}}, \Psi_{\overline{c}}, \Gamma_{\overline{c}})$ is the probability that the claimed speaker comes from a false identified emotion and a false identified gender".

The last step in the authentication procedure is to compete the "log-likelihood ratio" with the "threshold" ($\theta$) so as to admit or decline the requested speaker, *i.e.,*

Accept the claimed speaker if $\Lambda(O) \geq \theta$

Reject the claimed speaker if $\Lambda(O) < \theta$

Thresholding is often used to decide if a speaker is out of the set in open set speaker verification problems. Both types of error in speaker verification problem count on the threshold used in the decision making process. A firm value of threshold makes it harder for false speakers to be falsely accepted but at the cost of falsely rejecting true speakers. In contrast, a relaxed value of threshold eases true speakers to be accepted continuously at the expense of falsely accepting false speakers. To establish a suitable value of threshold that meets with a needed level of a true speaker rejection and a false speaker acceptance, it is essential to know the distribution of true speaker and false speaker scores. An acceptable method to build a reasonable value of threshold is to start with a relaxed initial value of threshold and then let it adjust by setting it to the average of the most fresh trial scores.

## 7.    Results and Discussion

In this study, a three-stage architecture has been proposed, executed, and tested to enhance the degraded "speaker verification performance in emotional environments". Our proposed framework has been tested on, based on HMMs (stage 1 and stage 3) and SPHMMs (stage 2) as classifiers, each of the in-house and EPST datasets. The weighting factor $\alpha$ in SPHMMs has been chosen to be equal 0.5 to keep away from biasing towards either acoustic or prosodic model.

In this work, stage 1 of the overall proposed framework yields 97.18% and 96.23% gender identification performance using the collected and EPST datasets, respectively. These two achieved performances are higher than those reported in some prior studies [33], [34]. Harb and Chen [33] attained 92.00% as gender identification performance in neutral environments. Vogt and Andre [34] obtained 90.26% as gender identification performance using Berlin German dataset.

The next stage is to recognize the unknown emotion of the claimed speaker given that his/her gender was recognized. This stage is called gender-dependent emotion identification problem. In this stage, SPHMMs has been used as classifier with $\alpha = 0.5$. Table 1 shows gender-dependent "emotion identification performance" based on SPHMMs using each of the in-house and EPST datasets. Based on this table, "average emotion identification performance" using the in-house and EPST datasets is 89.10% and 88.38%, respectively. These two values are higher than those reported in prior work by:

i) "Ververidis and Kotropoulos" [31] who attained 61.10% and 57.10% as "male and female average emotion identification performance", respectively.

ii) "Vogt and Andre" [34] who achieved 86.00% as gender-dependent "emotion identification performance" using Berlin dataset.

Table 1
Gender-dependent "emotion identification performance" using each of the in-house and EPST datasets

| Emotion | Emotion identification performance (%) | |
|---|---|---|
| | Collected dataset | EPST dataset |
| Neutral | 96.5 | 95.7 |
| Anger | 85.3 | 84.8 |
| Sadness | 86.1 | 86.0 |
| Happiness | 92.0 | 90.9 |
| Disgust | 85.8 | 85.1 |
| Fear | 88.9 | 87.8 |

Table 2 and Table 3 illustrate, respectively, male and female confusion matrices using the in-house dataset, while Table 4 and Table 5 demonstrate, respectively, male and female confusion matrices using EPST dataset. Based on these four matrices, the following general points can be noticed:

a. The most easily recognizable emotion is neutral, while the least easily recognizable emotions are anger/hot anger and disgust. Therefore, speaker verification performance is expected to be high when speakers speak neutrally without any emotion; on the other hand, the performance is predicted to be low when speakers talk angrily or disgustedly.

b. Column 3 "Anger" of Table 2, for example, states that 1% of the utterances that were uttered by male speakers in an anger emotion were assessed as generated in a neutral state. This column shows that anger emotion for male speakers has no confusion with happiness emotion (0%). This column also demonstrates that anger emotion for male speakers has the greatest confusion percentage with disgust emotion (4%). Hence, anger emotion is highly confusable with disgust emotion.

Table 2

Male "confusion matrix" of stage 2 of the three-stage architecture using the in-house dataset

| Emotion | "Percentage of confusion of the unknown emotion with the other emotions (%)" | | | | | |
|---|---|---|---|---|---|---|
| | "Neutral" | "Anger" | "Sadness" | "Happiness" | "Disgust" | "Fear" |
| "Neutral" | **99** | 1 | 1 | 1 | 0 | 0 |
| "Anger" | 0 | **90** | 1 | 1 | 3 | 3 |
| "Sadness" | 1 | 3 | **96** | 2 | 3 | 1 |
| "Happiness" | 0 | 0 | 0 | **94** | 1 | 1 |
| "Disgust" | 0 | 4 | 1 | 1 | **90** | 2 |
| "Fear" | 0 | 2 | 1 | 1 | 3 | **93** |

Table 3

Female "confusion matrix" of stage 2 of the three-stage architecture using the in-house dataset

| Emotion | "Percentage of confusion of the unknown emotion with the other emotions (%)" | | | | | |
|---|---|---|---|---|---|---|
| | "Neutral" | "Anger" | "Sadness" | "Happiness" | "Disgust" | "Fear" |
| "Neutral" | **99** | 1 | 1 | 1 | 0 | 0 |
| "Anger" | 0 | **91** | 1 | 1 | 2 | 2 |
| "Sadness" | 0 | 2 | **95** | 2 | 3 | 1 |
| "Happiness" | 1 | 0 | 1 | **95** | 1 | 1 |
| "Disgust" | 0 | 4 | 1 | 1 | **91** | 2 |
| "Fear" | 0 | 2 | 1 | 0 | 3 | **94** |

Table 4

Male "confusion matrix" of stage 2 of the three-stage architecture using EPST dataset

| Emotion | "Percentage of confusion of the unknown emotion with the other emotions (%)" | | | | | |
|---|---|---|---|---|---|---|
| | "Neutral" | "Hot Anger" | "Sadness" | "Happiness" | "Disgust" | "Panic" |
| "Neutral" | **99** | 4 | 1 | 1 | 1 | 1 |
| "Hot Anger" | 0 | **90** | 1 | 1 | 2 | 2 |
| "Sadness" | 0 | 1 | **96** | 1 | 3 | 1 |
| "Happiness" | 0 | 1 | 0 | **95** | 0 | 0 |
| "Disgust" | 0 | 3 | 1 | 1 | **90** | 2 |
| "Panic" | 1 | 1 | 1 | 1 | 4 | **94** |

Table 5

Female "confusion matrix" of stage 2 of the three-stage architecture using EPST dataset

| Emotion | "Percentage of confusion of the unknown emotion with the other emotions (%)" | | | | | |
|---------|-----------|-------------|-----------|--------------|-----------|---------|
| | "Neutral" | "Hot Anger" | "Sadness" | "Happiness" | "Disgust" | "Panic" |
| "Neutral" | **99** | 3 | 1 | 1 | 1 | 1 |
| "Hot Anger" | 0 | **91** | 1 | 1 | **2** | 1 |
| "Sadness" | 0 | 1 | **96** | 1 | 3 | 1 |
| "Happiness" | 1 | 1 | 0 | **96** | 0 | 0 |
| "Disgust" | 0 | **2** | 1 | 1 | **91** | **2** |
| "Panic" | 0 | **2** | 1 | 0 | 3 | **95** |

Table 6 gives percentage Equal Error Rate (EER) for speaker verification system in emotional environments based on the novel three-stage architecture in each of the collected and EPST datasets. The average value of percentage EER is 5.67% and 6.33% using the collected and EPST datasets, respectively. These values are less than those reported based on the two-stage framework proposed by Shahin [11]. This table shows that the least percentage EER takes place when speakers speak neutrally, while the greatest percentage EER occurs when speakers talk angrily or disgustedly. This table evidently yields higher percentage EER when speakers speak emotionally compared to when speakers speak neutrally. The reasons are accredited to:

Table 6

Percentage EER based on the three-stage architecture using the in-house and EPST datasets

| Emotion | EER (%) | |
|---------|-------------------|--------------|
| | Collected dataset | EPST dataset |
| Neutral | 1.0 | 1.5 |
| Anger/Hot Anger | 7.5 | 9.5 |
| Sadness | 6.0 | 6.5 |
| Happiness | 6.5 | 7.0 |
| Disgust | 7.0 | 8.0 |
| Fear/Panic | 6.0 | 5.5 |

1. Gender identification stage does not recognize the gender of the claimed speaker ideally. The average gender identification performance is 97.18% and 96.23% using the collected and EPST datasets, respectively.

2. Emotion identification stage is imperfect. The "average emotion identification performance" using the in-house and EPST datasets is 89.10% and 88.38%, respectively.

3. Speaker verification stage does not authenticate the claimed speaker perfectly. The average value of percentage EER is 5.67% and 6.33% using the collected and EPST datasets, respectively. The verification stage (stage 3) yields another system degradation performance in addition to the degradation in each of gender identification performance and emotion identification performance. This is because some claimants are rejected as imposters and some imposters are accepted as claimants. Consequently, the presented percentage EER in Table 6 is the resultant of percentage EER of each of stage 1, stage 2, and stage 3. The three-stage framework could have a negative impact on the overall speaker verification performance especially when both the gender (stage 1) and emotion (stage 2) of the claimed speaker has been falsely recognized.

In the current work, the attained average percentage EER based on the three-stage approach is less than that obtained in prior studies:

1) The author of [9] obtained 15.9% as an average percentage EER in emotional environments based on HMMs only.

2) The author of [11] achieved average percentage EER of 7.75% and 8.17% using the collected and EPST datasets, respectively.

3) The authors of [24] reported an average percentage EER of 11.48% in emotional

environments using GMM-UBM based on emotion-independent method.

Seven extensive experiments have been carried out in this research to test the achieved results based on the three-stage architecture. The seven experiments are:

(1)  Experiment 1: Percentage EER based on the proposed three-stage architecture has been competed with that based on the one-stage framework (gender-independent, emotion-independent, and text-independent speaker verification) using separately each of the collected and EPST datasets. Based on the one-stage approach and utilizing HMMs as classifiers, the percentage EER using the collected and EPST datasets is given in Table 7. This table gives percentage EER 14.75% and 14.58% using the collected and EPST datasets, respectively. It is apparent from Table 6 and Table 7 that the three-stage framework is superior to the one-stage approach.

Table 7
Percentage EER based on the one-stage approach using the
in-house and EPST datasets

| Emotion | EER (%) | |
|---|---|---|
| | Collected dataset | EPST dataset |
| Neutral | 6.0 | 6.0 |
| Angry/Hot Anger | 18.5 | 18 |
| Sad | 13.5 | 13.5 |
| Happy | 15.5 | 15.5 |
| Disgust | 16.5 | 16.5 |
| Fear/Panic | 18.5 | 18.0 |

To confirm whether EER differences (EER based on the three-stage framework and that based on the one-stage approach) are actual or just come from statistical variations, a "statistical significance test" has been conducted. The "statistical significance test" has been implemented based on the "Student's $t$ Distribution test".

In this work, $\overline{x}_{6,\text{collect}} = 5.67, \text{SD}_{6,\text{collect}} = 2.15, \overline{x}_{6,\text{EPST}} = 6.33, \text{SD}_{6,\text{collect}} = 2.49,$ $\overline{x}_{7,\text{collect}} = 14.75, \text{SD}_{7,\text{collect}} = 4.28, \overline{x}_{7,\text{EPST}} = 14.58, \text{SD}_{7,\text{EPST}} = 4.14$. These values have been computed based on Table 6 (collected and EPST datasets) and Table 7 (collected and EPST datasets), respectively. Based on these values, "the calculated $t$ value" using the collected dataset of both Table 6 and Table 7 is $t_{7,6\text{ (collected)}} = 2.119$ and the calculated $t$ value using EPST dataset of both Table 6 and Table 7 is $t_{7,6\text{ (EPST)}} = 1.896$. Each "calculated $t$ value" is greater than the "tabulated critical value at *0.05* significant level $t_{0.05} = 1.645$". Therefore, we can conclude based on this experiment that the three-stage speaker verification architecture is superior to the one-stage speaker verification framework. Hence, embedding both of gender and emotion identification stages into the one-stage speaker verification architecture in emotional environments significantly improves speaker verification performance compared to that without embedding these two stages. The conclusions of this experiment answer research question RQ1.1.

(2)    Experiment 2: Percentage EER based on the proposed three-stage framework has been competed with that based on the emotion-independent two-stage framework (gender-dependent, emotion-independent, and text-independent speaker verification) using independently each of the collected and EPST datasets. Based on this framework, the percentage EER based on the gender-dependent, emotion-independent, and text-independent approach and using the two speech datasets separately is illustrated in Table

8. This table yields average percentage EER of 11.67% and 11.92% using, respectively, the collected and EPST datasets.

Table 8

Percentage EER based on emotion-independent two-stage framework
using the in-house and EPST datasets

| Emotion | EER (%) | |
|---|---|---|
| | Collected dataset | EPST dataset |
| Neutral | 4.0 | 4.5 |
| Angry/Hot Anger | 14.5 | 15 |
| Sad | 10.5 | 11.5 |
| Happy | 12.0 | 12.5 |
| Disgust | 13.5 | 13.5 |
| Fear/Panic | 15.5 | 14.5 |

Using this table, $\bar{x}_{8,\text{collect}} = 11.67, SD_{8,\text{collect}} = 3.79, \bar{x}_{8,\text{EPST}} = 11.92, SD_{8,\text{collect}} = 3.52$, the calculated $t$ value, using the collected dataset of both Table 6 and Table 8, is $t_{8,6 \text{ (collected)}} = 2.007$ and the "calculated $t$ value", using EPST dataset of both Table 6 and Table 8, is $t_{8,6 \text{ (EPST)}} = 1.874$. Each "calculated $t$ value" is larger than the "tabulated critical value $t_{0.05} = 1.645$". Consequently, we can conclude based on this experiment that the three-stage speaker verification architecture outperforms the emotion-independent two-stage speaker verification framework. So, inserting "emotion identification stage" into the emotion-independent two-stage speaker verification architecture in emotional environments considerably enhances speaker verification performance compared to that without such a stage. In addition, the "calculated $t$ value", using the collected dataset of both Table 7 and Table 8, is $t_{8,7 \text{ (collected)}} = 1.961$ and the "calculated $t$ value", using EPST dataset of both Table 7 and Table 8, is $t_{8,7 \text{ (EPST)}} = 1.842$. Each "calculated $t$ value" is higher than the "tabulated critical value $t_{0.05} = 1.645$". Therefore, we can tell based on this experiment that the emotion-independent two-stage speaker verification architecture leads the one-stage speaker verification framework. So, adding emotion identification stage into the one-stage speaker verification architecture in emotional environments noticeably increases speaker verification performance compared to that without adding this stage. The conclusions of this experiment address research question RQ1.2.

(3)    Experiment 3: Percentage EER based on the proposed three-stage framework has been compared with that based on the gender-independent two-stage framework (gender-independent, emotion-dependent, and text-independent speaker verification) using individually each of the collected and EPST datasets. Based on this methodology, the attained percentage EER using the collected and EPST dataset is given in Table 9. This table gives average percentage EER of 7.75% and 8.17% using the collected and EPST datasets, respectively.

Table 9

Percentage EER based on gender-independent two-stage approach using the in-house and EPST datasets

| Emotion | EER (%) | |
|---|---|---|
| | Collected dataset | EPST dataset |
| Neutral | 1.5 | 2 |
| Angry/Hot Anger | 10.5 | 12 |
| Sad | 8 | 7.5 |
| Happy | 8.5 | 9 |
| Disgust | 9.5 | 10.5 |
| Fear/Panic | 8.5 | 8 |

Based on this table, the "calculated *t* value", using the collected dataset of both Table 6

and Table 9, is $t_{9,6 \text{ (collected)}} = 1.853$ and the "calculated $t$ value", using EPST dataset of both Table 6 and Table 9, is $t_{9,6 \text{ (EPST)}} = 1.792$. Each "calculated $t$ value" is greater than the "tabulated critical value $t_{0.05} = 1.645$". Therefore, we can infer, based on this experiment, that the three-stage speaker verification architecture is leader to the gender-independent two-stage speaker verification approach. Hence, adding gender identification stage into the gender-independent two-stage speaker verification architecture in emotional environments appreciably improves speaker verification performance compared to that without adding this stage. The conclusions of this experiment answer research question RQ1.3.

It is very important to make a comparison between Experiment 2 and Experiment 3 in terms of the performance. Since each one of these two experiments has two stages, it is very important to tell which two stages is more important than the other. Based on Table 8 and Table 9, the "calculated $t$ value" using the collected dataset $t_{9,8 \text{ (collected)}} = 1.749$ and the "calculated $t$ value" using EPST dataset $t_{9,8 \text{ (EPST)}} = 1.762$. It is evident from this experiment that emotion identification stage is more important than gender identification stage for speaker verification in emotional environments. Consequently, emotion information is more influential than gender information on speaker verification performance in these environments. However, merging and integrating gender information, emotion information, and speaker information into one system yields higher speaker verification performance than merging and integrating emotion information and speaker verification only into one system in such environments.

(4)    Experiment 4: As discussed earlier in this work, HMMs have been used as classifiers in stage 1 and stage 3, while SPHMMs have been used as classifiers in stage 2. In this experiment, the three-stage architecture has been assessed based on HMMs in all the three stages to compare the influence of using acoustic features with that using suprasegmental features on emotion identification (stage 2 of the three-stage architecture). In this experiment, Eq. (4) has become,

$$E^* = \arg \max_{m \geq e \geq 1} \left\{ P\left( O \middle| G^*, \lambda_E^e \right) \right\} \tag{10}$$

The achieved percentage EER based on this experiment is given in Table 10. This table yields 8.83% and 9.00% as average percentage EER using the collected and EPST datasets, respectively. To compete the impact between utilizing acoustic features and suprasegmental features on emotion identification stage in the novel three-stage framework, the "Student's $t$ Distribution test" has been performed on Table 6 and Table 10. The "calculated $t$ value" using the collected dataset is $t_{10,6 \text{ (collected)}} = 1.882$ and the "calculated $t$ value" using EPST dataset is $t_{10,6 \text{ (EPST)}} = 1.901$. Therefore, it is apparent from this experiment that using SPHHMs as classifiers in the emotion identification stage outperforms that using HMMs as classifiers in the same stage of the three-stage architecture. The conclusions of this experiment address research question RQ2.

Table 10

Percentage EER based on all HMMs three-stage architecture using the in-house and EPST datasets

| Emotion | EER (%) | |
|---|---|---|
| | Collected dataset | EPST dataset |
| Neutral | 2.0 | 1.5 |
| Anger/Hot Anger | 11.5 | 12.0 |
| Sadness | 9.5 | 10.0 |
| Happiness | 10.0 | 10.5 |
| Disgust | 11.0 | 10.5 |
| Fear/Panic | 9.0 | 9.5 |

Fig. 5 and Fig. 6 demonstrate Detection Error Trade-offs (DETs) curves using the collected and EPST datasets, respectively. Every curve compares speaker verification in emotional environments based on the three-stage framework with that based on each of one-stage, gender-dependent and emotion-independent, gender-independent and emotion-dependent, and all HMMs three-stage architectures in the same environments. These two figures apparently show that the three-stage architecture is superior to each one of these frameworks for speaker verification in such environments.

(5)    Experiment 5: The proposed three-stage architecture has been tested for diverse values of $\alpha$ (0.0, 0.1, 0.2,…, 0.9, 1.0). Fig. 7 and Fig. 8 illustrate average percentage EER based on the proposed framework versus the different values of $\alpha$ using the collected and EPST datasets, respectively. It is obvious from the two figures that as $\alpha$ increases, the average percentage EER decreases significantly and, consequently, increases speaker verification performance in emotional environments based on the three-stage framework except when speakers speak neutrally. The conclusion that can be made in this experiment is that SPHMMs have more impact than HMMs on speaker verification performance in these environments. The two figures also indicate that the least average percentage EER occurs when the classifiers are totally biased toward suprasegmental models ($\alpha = 1$) and no effect of the acoustic models ($\alpha = 0$).

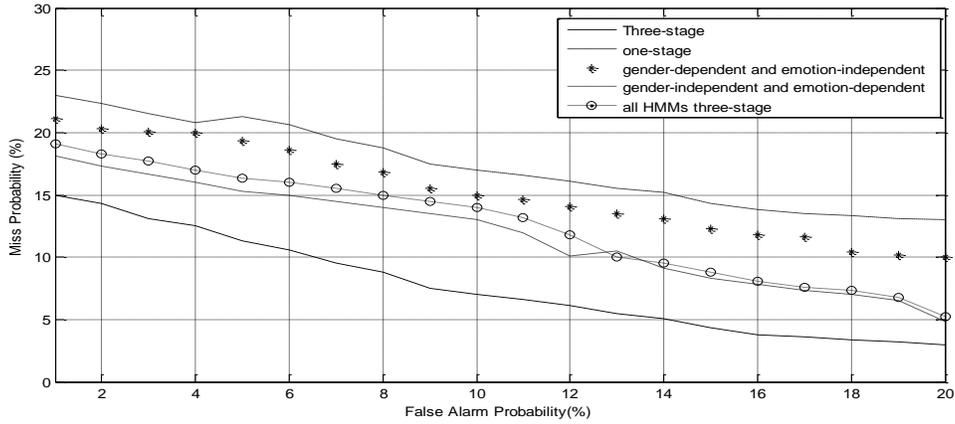

**Fig. 5.** DET curve based on each of three-stage, one-stage, gender-dependent and emotion-independent, gender-independent and emotion-dependent, and all HMMs three-stage architectures using the collected database

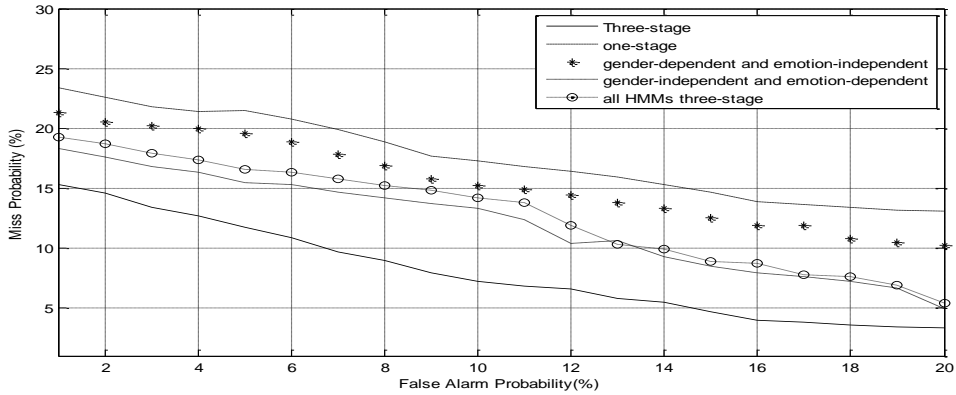

**Fig. 6.** DET curve based on each of three-stage, one-stage, gender-dependent and emotion-independent, gender-independent and emotion-dependent, and all HMMs three-stage architectures using EPST database

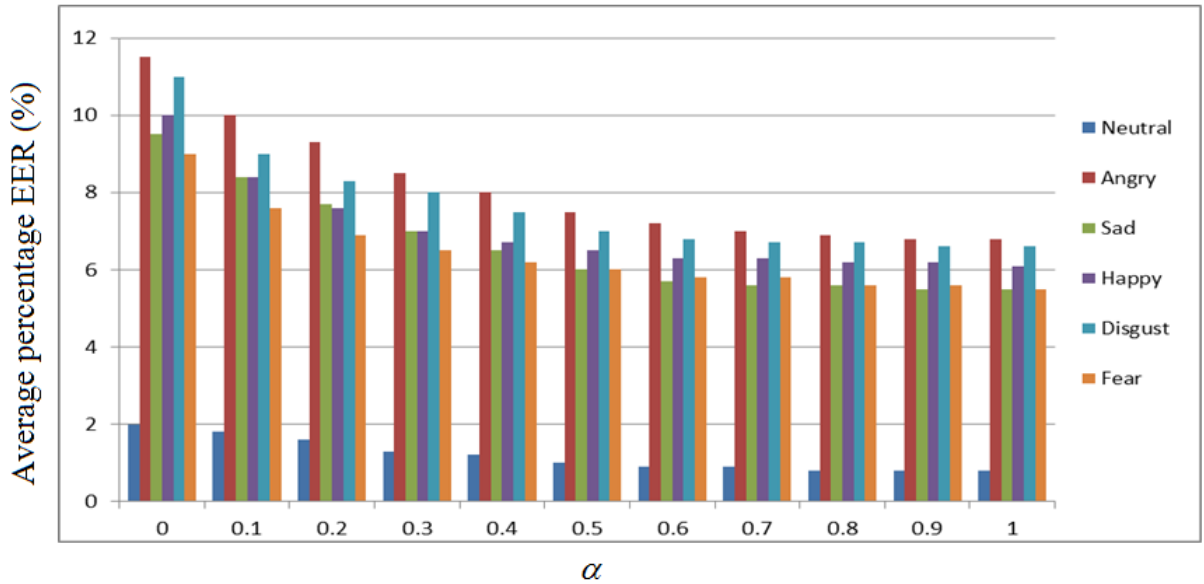

**Fig. 7**. Average percentage EER (%) versus $\alpha$ based on the three-stage framework using the collected database

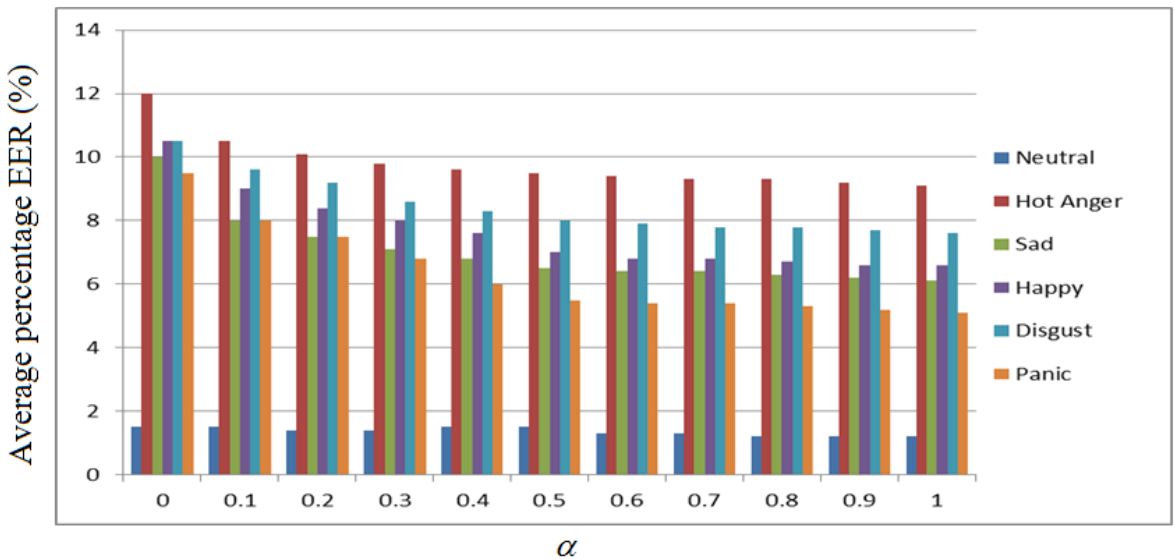

**Fig. 8**. Average percentage EER (%) versus $\alpha$ based on the three-stage framework using EPST database

(6)     Experiment 6: The novel three-stage architecture has been assessed for the "worst-case scenario". "Worst-case scenario" takes place when stage 3 gets untrue input from both the preceded two stages (stage 1 and stage 2). Hence, this scenario happens when speaker verification stage receives false identified gender and incorrect recognized emotion. The attained average percentage EER in the "worst-case scenario" based on SPHMMs when $\alpha = 0.5$ is 15.01% and 14.93% using the collected and EPST datasets, respectively. These achieved averages are very similar to those obtained using the one-stage approach (14.75% and 14.58% using the collected and EPST datasets, respectively).

(7)     Experiment 7: An "informal subjective assessment" of the proposed three-stage framework has been conducted with "ten (five male and five female") nonprofessional listeners (human judges)" using the collected speech dataset. These listeners were arbitrarily selected from distinct ages $(20 - 50$ years old). These judges were not used in collecting the collected speech dataset. A total of 960 utterances (20 speakers $\times$ 2 genders $\times$ 6 emotions $\times$ the last 4 sentences of the data corpus) have been utilized in this experiment. Each listener in this assessment is asked three sequential questions for every test utterance. The three successive questions are: identify the unknown gender of the claimed speaker, then identify the unknown emotion of the claimed speaker given his/her gender was recognized, and finally verify the claimed speaker provided both his/her gender and emotion were identified. Based on the subjective evaluation of this experiment, the average: gender identification performance, emotion identification performance, and speaker verification performance is 96.24%, 87.57%, and 84.37%, respectively. These averages are close to those achieved based on the novel three-stage speaker verification architecture.

# 8.    Concluding Remarks

In the present research, a novel three-stage speaker verification architecture has been introduced, executed, and tested to enhance "speaker verification performance in emotional environments". This architecture combines and integrates three sequential recognizers: gender identifier, followed by emotion identifier, followed by speaker verifier into one recognizer using both HMMs and SPHMMs as classifiers. This architecture has been assessed on two distinct and independent speech datasets: the in-house and EPST. Seven extensive experiments have been performed in this research to evaluate the proposed framework. Some conclusions can be drawn in this work. Firstly, speaker verification in emotional environments based on both gender cues and emotion cues is superior to each of that based on gender cues only, emotion cues only, and neither gender cues nor emotion cues. Secondly, as classifiers, SPHMMs outperform HMMs for speaker verification in these environments. The maximum average speaker verification performance takes place when the classifiers are entirely biased toward suprasegmental models and no impact of acoustic models. Thirdly, the three-stage framework functions nearly the same as the one-stage approach when the third stage of the three-stage architecture receives both an incorrect identified gender and a false identified emotion from the preceded two stages. Fourthly, emotion cues are more important than gender cues to speaker verification system. However, both of gender and emotion cues are more prominent than emotion cues only to speaker verification system in these environments. Finally, this study apparently shows that the emotional status of the claimed speaker has a negative influence on speaker verification performance.

In this work, two research questions: RQ1 and RQ2 were raised in "Section 2". Regarding RQ1.1, we showed in Experiment 1 of Section 7, that the proposed three-stage framework outperforms a single-stage framework. Moreover, Experiment 2 proves that the performance of the three-stage architecture is higher than the emotion independent two-stage approach and this answers the research question RQ1.2. To address research question RQ1.3, we conducted Experiment 3 and showed that the proposed three-stage model also surpasses the gender independent two-stage framework. Finally, in Experiment 4, we showed that the SPHMM classifier outperforms the HMM classifier, and this addresses RQ2.

Our proposed three-stage speaker verification approach has some limitations. First, in the three-stage framework, the required processing calculations and the time consumed are higher than those in the one-stage approach. Second, speaker verification performance based on the three-stage architecture is imperfect. This three-stage performance is the resultant of three non-ideal performances:

(a) The unknown gender of the claimed speaker is not 100% correctly identified in the first stage.

(b) The unknown emotion of the claimed speaker is imperfectly recognized in stage 2.

(c) The claimed speaker is non-ideally verified in the last stage.

For future work, our plan is to further alleviate speaker verification performance degradation in emotional environments based on proposing novel classifiers. Our plan also is to analytically work on the three-stage architecture to determine the performance of each stage individually and the overall performance of the three-stage speaker verification architecture; we intend to express the entire performance in terms of the performance of every stage.

**Acknowledgments**

The authors of this work would like to thank "University of Sharjah" for funding their work through the competitive research projects entitled "Emotion Recognition in each of Stressful and Emotional Talking Environments Using Artificial Models", No. 1602040348-P.

**"Conflict of Interest**: The authors declare that they have no competing interests".

**"Informed consent**: This study does not involve any animal participants".